\documentclass{article}
\usepackage[utf8]{inputenc}
\usepackage{ulem}
\usepackage{amssymb,amsmath}
\usepackage{graphicx,grffile}
\usepackage{caption}
\usepackage{subcaption}
\usepackage{xcolor}
\usepackage{xltabular}
\usepackage{listings}
\usepackage{authblk}
\usepackage{refcount}

\lstdefinestyle{fort}{language=Fortran,
  deletekeywords={dim,rank},
  basicstyle=\ttfamily,
  breaklines=true,
  keywordstyle=\color{blue},
  commentstyle=\color{green}\it,
  morecomment=[l][\color{red}]{!$}
}

    \makeatletter
\def\@fnsymbol#1{\ensuremath{\ifcase#1\or \dagger\or \ddagger\or
   \mathsection\or \mathparagraph\or \|\or **\or \dagger\dagger
   \or \ddagger\ddagger \else\@ctrerr\fi}}
    \makeatother

\newcommand{\vu}{\textbf{u}}

\newcommand{\vx}{\textbf{x}}

\newcommand{\vF}{\textbf{F}}

\newcommand{\ve}{\textbf{e}}
\newcommand{\vt}{\textbf{t}}

\newcommand{\oalpha}{\bar{\alpha}}

\title{IMEXLBM 1.0: A Proxy Application based on the Lattice Boltzmann Method for solving Computational Fluid Dynamic problems on GPUs}

\author[1]{Geng Liu\thanks{now at Argonne Leadership Computing Facility, Argonne National Laboratory, Lemont, IL, {email: gliu@anl.gov}}}
\author[2]{Saumil Patel\thanks{{email: spatel@anl.gov}}}
\author[2]{Ramesh Balakrishnan\thanks{{email: bramesh@anl.gov}}}
\author[1]{\\Taehun Lee\thanks{{email: thlee@ccny.cuny.edu}}}
\affil[1]{Department of Mechanical Engineering, The City College of the City University of New York, New York, NY}
\affil[2]{Computational Science Division, Argonne National Laboratory, Lemont, IL}
\date{January 2022}

\begin{document}
\maketitle
\section*{Abstract}
The US Department of Energy launched the Exascale Computing Project (ECP) in 2016 as part of a coordinated effort to achieve the next generation of high-performance computing (HPC) and to accelerate scientific discovery.
 The Exascale Proxy Applications Project began within the ECP to: (1) improve the quality of proxies created by the ECP (2) provide small, simplified codes which share important features of large applications and (3) capture programming methods and styles that drive requirements for compilers and other elements of the tool chain. This article describes one Proxy Application (or  ``proxy app") suite called IMEXLBM which is an open-source, self-contained code unit, with minimal dependencies, that is capable of running on heterogeneous platforms like those with graphic processing units (GPU) for accelerating the calculation. In particular, we demonstrate functionality by solving a benchmark problem in computational fluid dynamics (CFD) on the ThetaGPU machine at the Argonne Leadership Computing Facility (ALCF).  Our method makes use of a domain-decomposition technique in conjunction with the message-passing interface (MPI) standard for distributed memory systems. The OpenMP application programming interface (API) is employed for shared-memory multi-processing and offloading critical kernels to the device (i.e. GPU).  We also verify our effort by comparing data generated via CPU-only calculations with data generated with CPU+GPU calculations.  While we demonstrate efficacy for single-phase fluid problems, the code-unit is designed to be versatile and enable new physical models that can capture complex phenomena such as two-phase flow with interface capture.

\section{Introduction}
This project aims at Lattice Boltzmann Method (LBM) code development for heterogeneous platforms. LBM \cite{He1997} is a relatively novel approach to solve the Navier-Stokes equations (NSE) in the low-Mach number regime and its governing equation can be derived from the Boltzmann equation after discretizing the phase space with constant microscopic lattice velocities. One major drive behind the use of LBM in the CFD community is the ease of parallelization, but the increasing popularity of LBM can also be attributed to its unique feature: LBM solves a simplified Boltzmann equation that is essentially a set of 1st-order hyperbolic PDEs with constant microscopic lattice velocities, for which a plethora of simple yet excellent discretization schemes are available \cite{Patel2014}. Furthermore, all the complex non-linear effects are modeled locally at the grid points. Thus, the overall numerical scheme becomes extremely simple and efficient.  Furthermore, the linear advection part of LBM can be solved exactly due to unity CFL property. The absence of dispersive and dissipative errors in LBM makes it an ideal choice for the simulation of turbulent flows, interfacial flow, and acoustics, in which conservation of kinetic energy and isotropy play a crucial role.

The LBM has several features which are attractive to CFD researchers. In particular, it is explicit in time, has nearest neighbor communication, and the computational effort is in the collision step, which is localized at a grid point.  Data movement during the propagation (or streaming) step is a critical characteristic of the method and becomes the computational bottleneck during production runs. As such, the LBM is a memory bandwidth-bound scheme. Over the years, LB codes have been written and optimized for large clusters of commodity CPUs \cite{pohl2004performance}. In recent years, traditional CPUs have experienced larger and larger gaps in the relative speed between processor frequencies and memory transfer. This growing difference can lead to increased time-to-solution for memory-bandwidth bound approaches like LBM.  GPUs, however, have larger memory bandwidth rates as compared to traditional CPUs, and, can deliver more floating-point operations per second (FLOPs). As a result, a large effort has been underway to port and optimize the LBM for GPUs.  More recent work has focused on exploiting the parallelism of GPUs \cite{bailey2009accelerating,biferale2013optimized,calore2016massively,obrecht2013multi}.

The variety of heterogeneous architectures across vendors has presented HPC application developers with a major challenge. In particular, each architecture requires vendor-specific data structures, programming models, and implementations for the code to run efficiently. Therefore, developing the application for portability via vendor-specific implementations can be a time-consuming endeavor. To facilitate portability , in this work, we choose the OpenMP API \cite{OpenMP} which allows for annotation of the code through pragma directives that are placed around code sections (generally loops) for parallel execution. This approach leaves compilers to apply optimization steps that are specific to each target architecture.

In this work, we describe the structure and implementation of an open-source, self-contained code unit called \texttt{IMEXLBM} \cite{imexlb}, which uses the lattice Boltzmann method to solve problems in fluid dynamics.  Our code is written in Fortran 90 and uses the message passing interface (MPI) and OpenMP offloading directives for parallel computing on multiple processing cores (CPUs) and accelerator-devices (GPUs). The remainder of this paper is detailed as follows. In section \ref{lbm}, we provide a numerical description of the algorithm used in the LBM. Section \ref{description} provides details of the Fortran 90 code structure. This includes definitions and description of critical variables, subroutines, and OpenMP offloading directives. The code is capable of two and three-dimensional (2D \& 3D) simulations. LBM algorithms are included in several modules where relevant subroutines can be called to achieve parallel computations. This section also includes a flat profile to help us understand time-consumed by each subroutine. Section \ref{example} shows some qualitative results for benchmark problem in CFD. This is followed by section \ref{validation} which validates our implementation. Section \ref{performance} outlines some initial and preliminary performance numbers. A summary of our conclusions can be found in section \ref{conclusion}

\section{Lattice Boltzmann Method}\label{lbm}
\subsection{Governing Equation}
\begin{figure}
\centering
  \begin{subfigure}[b]{0.3\textwidth}
    \centering
    \includegraphics[width=\textwidth]{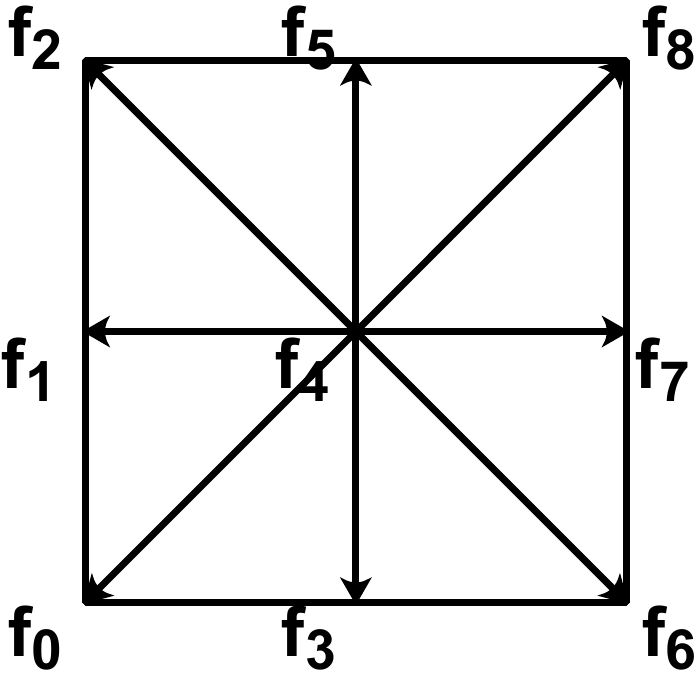}
    \caption{D2Q9 DVM.}
  \end{subfigure}
  \hfil
  \begin{subfigure}[b]{0.3\textwidth}
    \centering
    \includegraphics[width=\textwidth]{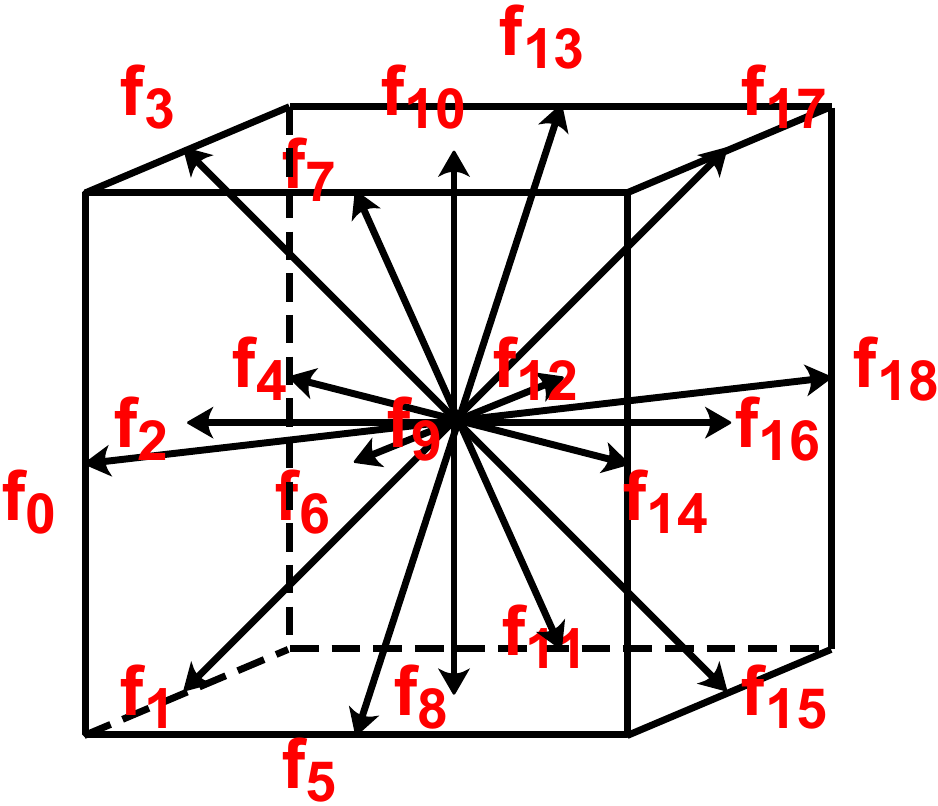}
    \caption{D3Q19 DVM.}
  \end{subfigure}
  \hfil
  \begin{subfigure}[b]{0.3\textwidth}
    \centering
    \includegraphics[width=\textwidth]{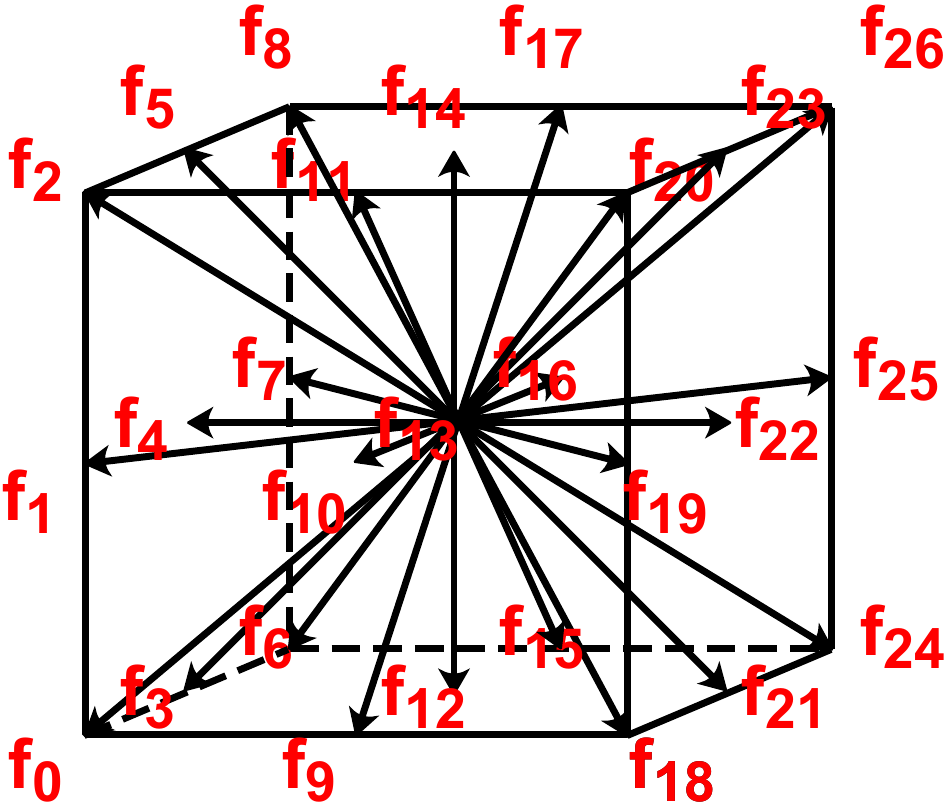}
    \caption{D3Q27 DVM.}
  \end{subfigure}
  \caption{2D and 3D DVMs. \label{DdQm}}
\end{figure}

The discrete Boltzmann equation (DBE) with the Bhatnagar, Gross and Krook (BGK) \cite{Bhatnagar1954} collision operator and  is given by
\begin{eqnarray}
\frac{\partial f_{\alpha}}{\partial t}+\ve_{\alpha}\cdot\nabla f_{\alpha}=-\frac{f_{\alpha}-f^{eq}_{\alpha}}{\tau},~~(\alpha=0,1,...,m-1),\label{DBE}
\end{eqnarray}
where $f_{\alpha}$ is the particle distribution function (PDF) in the $\alpha$ direction, $t$ is time, $\ve_{\alpha}$ is the corresponding microscopic particle velocity (or lattice velocity), $\tau$ is the relaxation time, and $m$ is the number of lattice velocities. The equilibrium distribution function $f^{eq}_{\alpha}$ is given by
\begin{eqnarray}
f^{eq}_{\alpha}=t_{\alpha}\rho\left[1+\frac{\ve_{\alpha}\cdot\vu}{c_s^2}+\frac{\left(\ve_{\alpha}\cdot\vu\right)^2}{2c_s^4}
-\frac{\vu\cdot\vu}{2c_s^2}\right],\label{feq}
\end{eqnarray}
where $t_{\alpha}$ is the weighting factor \cite{Qian2000}, $\rho$ is the macroscopic density, $\vu$ is the macroscopic velocity, and $c_s$ is the speed of sound. Qian  et al. (1992) has proposed a D$d$Q$m$ discrete velocity model (DVM), where $d$ represents the dimension, and $m$ represents the number of lattice velocities \cite{Qian1992}. Fig. \ref{DdQm} shows the DVMs of D2Q9, D3Q19 and D3Q27.
The numbering of the PDFs is a choice of developer and does not affect the LBM algorithm. In this work it is chosen as shown in Fig. \ref{DdQm}. These orders of lattice velocity directions are designed such that it is easy to find $\oalpha=m-1-\alpha$, which is the opposite direction of $\alpha$. This arrangement is especially useful when bounce back boundary condition is considered. Taking D2Q9 as an example, the configuration of DVM is
\begin{eqnarray}
&\ve=
\begin{bmatrix}
-1&-1&-1& 0& 0& 0& 1& 1& 1\\
-1& 0& 1&-1& 0& 1&-1& 0& 1
\end{bmatrix}c,\\
& \vt=
\begin{bmatrix}
\displaystyle\frac{1}{36}&\displaystyle\frac{1}{9}&\displaystyle\frac{1}{36}&\displaystyle\frac{1}{9}&\displaystyle\frac{4}{9}&
\displaystyle\frac{1}{9}&\displaystyle\frac{1}{36}&\displaystyle\frac{1}{9}&\displaystyle\frac{1}{36}
\end{bmatrix},
\end{eqnarray}
where $c$ is the ratio of grid size and time increment, and normally appears as 1.

The original equilibrium distribution of LBM is designed for weakly compressible fluid. An incompressible LBM model is carried out here with the equilibrium distributions expressed by
\begin{eqnarray}
\hat{f}^{eq}_{\alpha}=f^{eq}_{\alpha}-t_{\alpha}\left(\rho-\frac{p}{c_s^2}\right)=t_{\alpha}\left[\frac{p}{c_s^2}+\rho_0\left(\frac{\ve_{\alpha}\cdot\vu}{c_s^2}+\frac{\left(\ve_{\alpha}\cdot\vu\right)^2}{2c_s^4}
-\frac{\vu\cdot\vu}{2c_s^2}\right)\right],\label{incompressible_feq}
\end{eqnarray}
where $\rho_0$ is constant.
Substituting $\hat{f}_{\alpha}$ into fully discretized DBE (Eq. (\ref{DBE})) will generate the incompressible LBM model \cite{He19971,Lin1996,Zou1995}. For convenience, $\hat{f}_{\alpha}$ is denoted by $f_{\alpha}$ hereafter. The evolution of LBM is normally implemented with two major steps, propagation and collision, in their general forms,
\begin{eqnarray}
\textnormal{collision:}&&f_{\alpha}^{\ast}=f_{\alpha}-\frac{1}{\frac{\tau}{\delta_t}+\frac{1}{2}}(f_{\alpha}-f_{\alpha}^{\text{eq}}),\label{collision}\\
\textnormal{propagation:}&&f_{\alpha}(\vx,t)=f^{\ast}_{\alpha}(\vx-\ve_{\alpha} \delta_t,t-\delta_t)\textnormal{~~~~in~}\Omega \textnormal{~for~}\alpha=0,1,...,m-1,\nonumber\\\label{propagation}
\end{eqnarray}
where the macroscopic pressure and velocity are recovered by
\begin{eqnarray}
&\displaystyle p=c_s^2\sum_{\alpha}f_{\alpha},\\
&\displaystyle\vu=\frac{1}{\rho}\sum_{\alpha}\ve_{\alpha}f_{\alpha}.
\end{eqnarray}

\subsection{Auxiliary Conditions}
There are many treatments to the initial and boundary conditions of LBM. In this work, the PDFs are initialized by the equilibrium values obtained from given velocity and pressure fields.

The motionless no-slip boundaries are modeled by the second-order modified nodal bounce back \cite{Ziegler1993}. In this scheme, outgoing PDFs are replicated in their opposite directions after hitting the wall. A collision operation is then applied to the boundary nodes like in the fluid bulk. In the next propagation step the updated incoming PDFs are streamed back to the fluid nodes. This process can be expressed by
\begin{eqnarray}
f_{\alpha}=f_{\oalpha},\label{bb}
\end{eqnarray}
where $f_{\oalpha}$ and $f_{\alpha}$ are in opposite directions.

\section{Description of Code}\label{description}

This program contains three modules: the MPI setup module, the Physics setup module, and the LBM module. 2D and 3D modules are different and thus written in separate files. Subroutines in the modules are called in the main code.

\begin{figure}
\centering
\includegraphics[width=\textwidth]{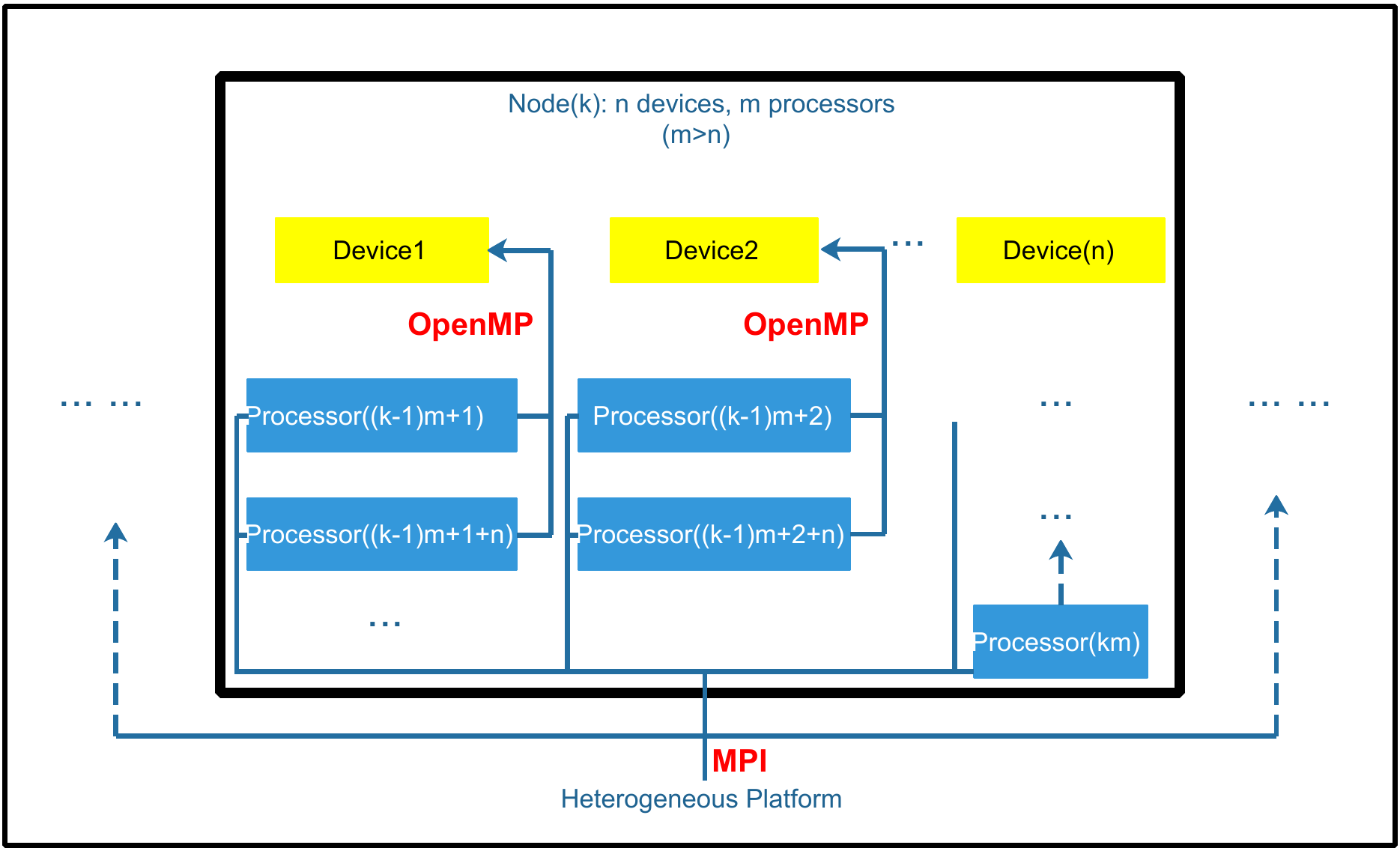}
\caption{The relationship between the general structure of this program and the heterogeneous platform.}\label{structure}
\end{figure}

The computing tasks are assigned to multiple processors through MPI setup in the base level. On top of that, each processor can offload its task to one accelerating device with OpenMP directives. The relationship between the general structure of this program and the heterogeneous platform is shown in Fig. \ref{structure}.

Three module files and one main code file can form the code package. The modules are CART\_MPI, INITIALIZATION and LBM. They are written in \texttt{cart\_mpi.f90}, \texttt{initialization.f90} and \texttt{lbm.f90} respectively for 2D cases, and \texttt{cart\_mpi\_3d.f90}, \texttt{initialization\_3d.f90} and \texttt{lbm\_3d.f90} respectively for 3D cases. Hereafter the main code, modules, variables and subroutines are explained.

\subsection{Main Code}
In the main code, module LBM is included. The code starts with initializing MPI and obtaining MPI rank and number of processors. Then it calls \texttt{DataInput} subroutine to read the physical and geometry setup. The \texttt{MPISetup} subroutine is called to 2D or 3D cartesian MPI. The lattice model (DVM) is defined by the \texttt{InitLattice} subrtoutine. Arrays for PDFs and other fields are allocated in \texttt{AllocateArrays} subroutine. The geometry is defined by \texttt{SetGeometry} subroutine. Users can modify \texttt{SetGeometry} and \texttt{DataInput} to build up their own problems.

Before starting the LBM algorithm, the different default device (GPU) numbers (variable \texttt{gpuid}) are assigned to each MPI rank, and the allocated arrays are offloaded to devices through the OpenMP directive !\$OMP TARGET ENTER DATA. The meaning of the arrays will be explained in the descriptions modules. The device related lines should be commented when OpenMP target offloading is turned off.

The \texttt{InitUP} and \texttt{InitPDF} subroutines initializes velocity, pressure and PDF fields. The variable \texttt{count}, initialized as 0, is introduced to manage output files. This variable accumulates with each output.

The main loop goes from 0 to maximum number of steps. Inside the loop, the index \texttt{iter} is checked in the first place. Every \texttt{interv} steps, user required runtime values can be printed and the simulation results can be written to defined files. The subroutine \texttt{WriteBinary} uses MPI writing to generate a single binary result file in PLT format. This file can be read by softwares like Tecplot and Paraview. Details of the subroutines are explained in the modules.

The \texttt{Collision}, \texttt{BoundaryCondition}, \texttt{Propagation} and \texttt{PostProcessing} subroutines construct the main algorithm of LBM. The MPI communication happens in the beginning of \texttt{BoundaryCondition} subroutine. For open boundaries, part of the boundary conditions are coded in the \texttt{PostProcessing} subroutine, but the main function of \texttt{PostProcessing} is to recover macroscopic fields by taking the moments of PDFs.

Once the job is done, the directive !\$OMP TARGET EXIT DATA terminates the OpenMP process and deletes the device variables. The CPU memories are also released with the subroutine \texttt{DeAllocateArrays}. In the end, MPI is terminated to finalize the program.

\subsection{Module CART\_MPI}
This module generates a cartesian MPI communicator, and defines the arrays distributed to each MPI processor. The module also defines the data types for sending and receiving subarrays. MPI library is included in this module.

\subsubsection{Variables}

\begin{xltabular}{\linewidth}{ l | X }
  \caption{Description of Variables used in Module CART\_MPI}
 \label{table: var_description_cart_mpi}\\
 \hline \hline

\textbf{\normalsize Variable} & \textbf{\normalsize Definition}  \\
 \hline
\endfirsthead
 \hline \hline

\textbf{\normalsize Variable} & \textbf{\normalsize Definition}  \\
 \hline
\endhead

\textbf{\texttt{dim}} & Dimension (2 or 3) of the problem. \\ \hline

\textbf{\texttt{nq}} & Number of lattice velocities.\\ \hline

\textbf{\texttt{ghost}} & Thickness of ghost nodes. Ghost nodes are extensions of arrays prepared for communication and boundary conditions. \\ \hline

\textbf{\texttt{num\_neighbots}} & Number of for each MPI processor.\\ \hline

\textbf{\texttt{nx}} & Number of points in x direction.\\ \hline

\textbf{\texttt{ny}} & Number of points in y direction.\\ \hline

\textbf{\texttt{nz}} & Number of points in z direction.\\ \hline

\textbf{\texttt{ierr}} & MPI error signal.\\ \hline

\textbf{\texttt{num\_procs}} & Number of MPI processors.\\ \hline

\textbf{\texttt{rank}} & 1D ID of this MPI processor.\\ \hline

\textbf{\texttt{cart\_coords}} & Cartesian coordinates of current MPI processor.\\ \hline

\textbf{\texttt{cart\_num\_procs}} & Number of MPI processors in different cartesian directions.\\ \hline

\textbf{\texttt{cart\_periods}} & Flags for periodic communication in each direction.\\ \hline

\textbf{\texttt{global\_length}} & Directional length of whole domain.\\ \hline

\textbf{\texttt{local\_length}} & Directional length of array distributed to current MPI processor.\\ \hline

\textbf{\texttt{local\_start}} & Local starting point's position in the whole domain.\\ \hline

\textbf{\texttt{local\_end}} & Local ending point's position in the whole domain.\\ \hline

\textbf{\texttt{array\_size}} & Size of scalor field array in current processor (reshaped to 1D).\\ \hline

\textbf{\texttt{neighbor\_procs}} & ID of each neighbor MPI processor.\\ \hline

\textbf{\texttt{X\_RECV\_TYPE}}\footnote{\label{X}X: INT-integer, F-PDF (or variable of same dimension), U-Velocity (or variable of same dimension), DPR-double precision, LOG-logical.} & Subarray datatypes for receiving data from each neighbor processor.\\ \hline

\textbf{\texttt{X\_SEND\_TYPE}}\footnotemark[\getrefnumber{X}] & Subarray datatypes for sending data from each neighbor processor.\\ \hline

\end{xltabular}

\subsubsection{Subroutines}
\begin{xltabular}{\linewidth}{ l | X }
  \caption{Description of Subroutines used in Module CART\_MPI}
 \label{table: sub_description_cart_mpi}\\
 \hline \hline

\textbf{\normalsize Subroutine} & \textbf{\normalsize Definition}  \\
 \hline
\endfirsthead
 \hline \hline

\textbf{\normalsize Subroutine} & \textbf{\normalsize Definition}  \\
 \hline
\endhead

\textbf{\texttt{MPISetup}} & By calling other subroutines, this subroutine generates a cartesian MPI communicator, and defines the arrays distributed to each MPI processor as well as the data types for sending and receiving subarrays. This subroutine is a package of subroutines SetCartesian, SetLocalLength and SetNeighbors. \\ \hline

\textbf{\texttt{SetCartesian}} & This subroutine generates a cartesian MPI communicator. It optimizes the number of processors in each direction.\\ \hline

\textbf{\texttt{SetLocalLength}} & Based on the communicator generated in SetCartesian subroutine, this subroutine generates the local array information, including the local array length in each direction, and the starting and ending points' positions in the whole array. \\ \hline

\textbf{\texttt{SetNeighbors}} & This subroutine defines the neighbors of this MPI processor, and generates sending and receiving subarray datatypes for communication with these neighbors.\\ \hline

\textbf{\texttt{PassF}}& This subroutine is called when communication of the array (PDF or variable of the same dimension) is needed. The communication is based on the sending and receiving subarray datatypes generated in SetNeighbors. For arrays of other data types, PassInt, PassD, PassU and PassLog can also be defined similarly.\\ \hline

\end{xltabular}

\subsection{Module INITIALIZATION}
This module includes the variables and subroutines for fluid initialization. CART\_MPI module is used in this module.
\subsubsection{Variables}

\begin{xltabular}{\linewidth}{ l | X }
  \caption{Description of Variables used in Module INITIALIZATION}
 \label{table: var_description_init}\\
 \hline \hline

\textbf{\normalsize Variable} & \textbf{\normalsize Definition}  \\
 \hline
\endfirsthead
 \hline \hline

\textbf{\normalsize Variable} & \textbf{\normalsize Definition}  \\
 \hline
\endhead

\textbf{\texttt{charlength}} & Characteristic length. \\ \hline

\textbf{\texttt{pos}} & Case (cylinder flow) related parameter: position of cylinder in x-direction.\\ \hline

\textbf{\texttt{re}} & Reynolds number. \\ \hline

\textbf{\texttt{uu}} & Characteristic velocity magnitude. \\ \hline

\textbf{\texttt{ma}} & Mach number. \\ \hline

\textbf{\texttt{radius}} & Case (cylinder flow) related parameter: cylinder radius, half of \texttt{charlength}. \\ \hline

\textbf{\texttt{nu}} & Kinematic viscosity in lattice units. \\ \hline

\textbf{\texttt{rho0}} & Fluid density, constant in incompressible model, 1. \\ \hline

\textbf{\texttt{dt}} & Time increment in lattice unit, 1. \\ \hline

\textbf{\texttt{ome}} & Relaxation parameter, 1/(tau+0.5). \\ \hline

\textbf{\texttt{tau}} & Relaxation time. \\ \hline

\textbf{\texttt{t\_intv}} & Dimensionless time increment. \\ \hline

\textbf{\texttt{inter}} & Time interval size (dimensionless) for monitor or output. \\ \hline

\textbf{\texttt{max\_t}} & Maximum time (dimensionless). \\ \hline

\textbf{\texttt{max\_step}} & Maximum number of time steps. \\ \hline

\textbf{\texttt{iter}} & Iteration index. \\ \hline

\textbf{\texttt{interv}} & Time interval size in lattice units for monitor or output. \\ \hline

\textbf{\texttt{count}} & This variable is initialized as 0, and is introduced to manage output files. It accumulates with each output. \\ \hline

\textbf{\texttt{t}} & Lattice weights. \\ \hline
\textbf{\texttt{e}} & Lattice velocities. \\ \hline
\textbf{\texttt{f, fb}} & PDF and backup PDF arrays. \\ \hline

\textbf{\texttt{p}} & Pressure array. \\ \hline

\textbf{\texttt{geo}} & Geometry array. \\ \hline

\textbf{\texttt{bx}}\footnote{\label{x}x: l-left, r-right, u-top, d-bottom, f-front, b-back, \_user/user-user defined.} & Vectors that store boundary points' indices. \\ \hline

\textbf{\texttt{fluid\_id}} & Vector that store fluid bulk points' indices. \\ \hline

\textbf{\texttt{dy}} & The index increase in 1D arrays reshaped from 3D arrays per 1 y-coordinate increment. \\ \hline

\textbf{\texttt{dz}} & The index increase in 1D arrays reshaped from 3D arrays per 1 z-coordinate increment. \\ \hline

\textbf{\texttt{size\_fluid}} & Size of \texttt{fluid\_id}. \\ \hline

\textbf{\texttt{x\_size}}\footnotemark[\getrefnumber{x}] & Size of \texttt{bx}. \\ \hline

\textbf{\texttt{u}} & Fluid velocity array. \\ \hline

\end{xltabular}

\subsubsection{Subroutines}

\begin{xltabular}{\linewidth}{ l | X }
  \caption{Description of Subroutines used in Module INITIALIZATION}
 \label{table: sub_description_init}\\
 \hline \hline

\textbf{\normalsize Subroutine} & \textbf{\normalsize Definition}  \\
 \hline
\endfirsthead
 \hline \hline

\textbf{\normalsize Subroutine} & \textbf{\normalsize Definition}  \\
 \hline
\endhead

\textbf{\texttt{DataInput}} & This subroutine reads data from \texttt{input.in} file, and generates derived initial parameters. \\ \hline

\textbf{\texttt{InitLattice}} & This subroutine defines the lattice velocities and weights, and can be modified to satisfy user's requirement for DVM.\\ \hline

\textbf{\texttt{AllocateArrays}} & This subroutine allocates local arrays on each MPI processor. \\ \hline

\textbf{\texttt{DeAllocateArrays}} & This subroutine deallocates local arrays.\\ \hline

\textbf{\texttt{SetGeometry}}& This subroutine initializes the geometry according to the setup of the simulation. \texttt{geo} is a logical array where fluid nodes are represented by 1 and solid nodes are represented by 0. The indices for boundary nodes and fluid bulk nodes are also generated in this subroutine.\\ \hline

\textbf{\texttt{InitUP}} & This subroutine  can be offloaded to device by OpenMP directives. It initializes the fluid's velocity and pressure fields and/or other fields. The initialization can be modified according to user's requirements.\\ \hline

\end{xltabular}

\subsection{Module LBM}
This module includes the subroutines for LBM algorithms. INITIALIZATION module is used in this module.
\subsubsection{Variables}

\begin{xltabular}{\linewidth}{ l | X }
  \caption{Description of Variables used in Module LBM}
 \label{table: var_description_lbm}\\
 \hline \hline

\textbf{\normalsize Variable} & \textbf{\normalsize Definition}  \\
 \hline
\endfirsthead
 \hline \hline

\textbf{\normalsize Variable} & \textbf{\normalsize Definition}  \\
 \hline
\endhead

\textbf{\texttt{um}} & Maximum velocity magnitude on current processor. \\ \hline

\textbf{\texttt{offset}} &Integer of MPI\_OFFSET\_KIND used in MPI output. \\ \hline

\textbf{\texttt{idn}} & Index for indices arrays. \\ \hline

\textbf{\texttt{id}} & Index for field arrays. \\ \hline

\textbf{\texttt{iq}} & Index for lattice directions. \\ \hline

\textbf{\texttt{io}} & Index for lattice direction opposite to \texttt{iq}. \\ \hline

\textbf{\texttt{ind}} & Index for dimensions. \\ \hline

\textbf{\texttt{udu}} & Velocity squared. \\ \hline

\textbf{\texttt{edu}} & Inner product of physical velocity and lattice velocity. \\ \hline

\end{xltabular}

\subsubsection{Subroutines}
\begin{xltabular}{\linewidth}{ l | X }
  \caption{Description of Subroutines used in Module LBM}
 \label{table: sub_description_lbm}\\
 \hline \hline

\textbf{\normalsize Subroutine} & \textbf{\normalsize Definition}  \\
 \hline
\endfirsthead
 \hline \hline

\textbf{\normalsize Subroutine} & \textbf{\normalsize Definition}  \\
 \hline
\endhead

\textbf{\texttt{InitPDF}} & This subroutine can be offloaded to device by OpenMP directives, it initializes and PDFs, it should be modified when equilibrium PDF is changed. \\ \hline

\textbf{\texttt{Collision}} & This subroutine can be offloaded to device by OpenMP directives, it is the collision process of LBM. The collision result of array \texttt{f} is stored in array \texttt{fb}, it should be modified as collision model changes. \\ \hline

\textbf{\texttt{Propagation}} & This subroutine can be offloaded to device by OpenMP directives, it is the propagation process of LBM. The propagation result of array \texttt{fb} is stored in array \texttt{f}. \\ \hline

\textbf{\texttt{BoundaryCondition}} & This subroutine can be offloaded to device by OpenMP directives, it applies the boundary conditions, including no-slip boundary condition (bounce back) and open boundary conditions. Before non-local operations communication should be applied. This condition is a pre-propagation condition, post-propagation conditions should be transformed to equivalent pre-propagation conditions, or be applied in the PostProcessing subroutine. \\ \hline

\textbf{\texttt{PostProcessing}} & This subroutine can be offloaded to device by OpenMP directives, it evaluates the physical properties, including the pressure and velocities, by taking the moments of PDFs. Some post-propagation boundary conditions are also included here. \\ \hline

\textbf{\texttt{Monitor}} & This subroutine prints the global maximum magnitude of velocity or optionally other fluid properties. The local part of this subroutine is offloaded to devices, and the global part and printing is done on CPU processors. \\ \hline

\textbf{\texttt{WriteBinary}} & This subroutine writes the simulation result to binary PLT output files. It contains function called str2ascii, which converts a string to ASCII codes. \\ \hline

\end{xltabular}

\subsection{Code Profile}

A test simulation of flow past a 3D sphere is performed to generate a time-profile of the code. The size of the geometry is 512 by 128 by 128 with detailed descriptions of the problem found in Sections \ref{validation} and \ref{performance}.
A flat profile of the code was generated using this simulation. The code was run for $6401$ timesteps, using 8 MPI ranks, with each rank bound to a single GPU on ThetaGPU, therefore using all 8 GPUs of the node. The breakdown of time-spent per kernel is shown below:
\begin{xltabular}{\linewidth}{m{0.05\textwidth}m{0.06\textwidth}m{0.05\textwidth}m{0.05\textwidth}m{0.05\textwidth}m{0.05\textwidth}m{0.4\textwidth}}
  \caption{The breakdown of time-spent per kernel}\footnote{\% time: the percentage of the total running time of the
program used by this function.\\
cumulative seconds: a running sum of the number of seconds accounted
 for by this function and those listed above it.\\
self  seconds:  the number of seconds accounted for by this
function alone.  This is the major sort for this listing.\\
calls:   the number of times this function was invoked, if
           this function is profiled, else blank.\\
self ms/call:   the average number of milliseconds spent in this
function per call, if this function is profiled, else blank.\\
total ms/call:  the average number of milliseconds spent in this
function and its descendents per call, if this function is profiled, else blank.\\
name:   the name of the function.  This is the minor sort
           for this listing. The index shows the location of
	   the function in the gprof listing. If the index is
	   in parenthesis it shows where it would appear in
	   the gprof listing if it were to be printed.}
 \label{breakdown}\\
 \hline \hline

\textbf{\footnotesize \% time} & \textbf{\footnotesize cumul- ative seconds}& \textbf{\footnotesize self seconds}& \textbf{\footnotesize calls}& \textbf{\footnotesize self ms/call}& \textbf{\footnotesize total ms/call}& \textbf{\footnotesize name}  \\
 \hline
\endfirsthead
 \hline \hline

\textbf{\footnotesize \% time} & \textbf{\footnotesize cumul- ative seconds}& \textbf{\footnotesize self seconds}& \textbf{\footnotesize calls}& \textbf{\footnotesize self ms/call}& \textbf{\footnotesize total ms/call}& \textbf{\footnotesize name}  \\
 \hline
\endhead

  0.00    &  0.11   &  0.00  &      1   &  0.00  & 100.01 & MAIN\\\hline
 45.46   &   0.05  &   0.05   &  6401&  0.01  &   0.01& lbm: boundarycondition\\ \hline
 18.18  &    0.07   &  0.02  &   6401  &   0.00  &   0.00 & lbm: collision\\ \hline
  9.09  &    0.08   &  0.01  &   6401  &   0.00  &   0.00 & lbm: postprocessing\\\hline
  9.09   &   0.10   &  0.01  &      1  &  10.00  &  10.00 & lbm: writebinary\\\hline
  0.00    &  0.11   &  0.00  &   6401   &  0.00  &   0.00 & lbm: propagation\\\hline
  0.00    &  0.11   &  0.00  &     11   &  0.00  &   0.00 & lbm: monitor\\\hline
  0.00    &  0.11  &   0.00   &     1   &  0.00  &   0.00 & lbm: initpdf\\\hline
  9.09    &  0.11   &  0.01  &          &        &        & nv lbm: collision, F1L56 3, F1L56 4\\\hline
  9.09  &    0.09   &  0.01  &      1  &  10.00  &  10.00 & init: setgeometry\\\hline
  0.00    &  0.11   &  0.00   &     1   &  0.00  &   0.00 & init: allocatearrays\\\hline
 0.00    &  0.11   &  0.00   &     1   &  0.00  &   0.00 & init: datainput\\\hline
 0.00    &  0.11   &  0.00   &     1   &  0.00  &   0.00 & init: deallocatearrays\\\hline
 0.00    &  0.11   &  0.00   &     1   &  0.00  &   0.00 & init: initlattice\\\hline
  0.00    &  0.11   &  0.00   &     1   &  0.00  &   0.00 &init: initup\\\hline
  0.00     & 0.11   &  0.00  &   6401   &  0.00  &   0.00 & cart\_mpi: passf\\\hline
  0.00    &  0.11   &  0.00  &      1   &  0.00  &   0.00 & cart\_mpi: mpisetup\\\hline
  0.00    &  0.11   &  0.00  &      1   &  0.00  &   0.00 & cart\_mpi: passint\\\hline
  0.00    &  0.11   &  0.00  &      1   &  0.00  &   0.00 & cart\_mpi: setcartesian\\\hline
  0.00    &  0.11   &  0.00  &      1   &  0.00  &   0.00 & cart\_mpi: setlocallength\\\hline
  0.00    &  0.11   &  0.00   &     1   &  0.00  &   0.00 & cart\_mpi: setneighbors\\\hline

\end{xltabular}

According to Tab. \ref{breakdown}, \texttt{BoundaryCondition} is the most time-consuming subroutine.  This is because the communication is in this subroutine. PDFs being copied to CPU memory to process MPI communication is probably not an efficient way. Device to device communication should be considered in future versions.

\texttt{Collision} is another time-consuming subroutine, it takes around 18\% of the total time, about twice as the time used for post-processing. The executing time for \texttt{Propagation} subroutine is negligible compared to the other subroutines.

\subsection{OpenMP Offloading Directives}
Most subroutines in LBM module can be offloaded to devices for acceleration. These offloading processes are achieved by making use of OpenMP directives. The \texttt{Collision} subroutine (Eq. (\ref{collision})) is shown here as examples to explain the mechanism.
\begin{lstlisting}[style=fort]
subroutine Collision
    integer idn,id,iq,ind
    double precision udu,edu,feq

    ! OpenMP offloading directive
    !$OMP TARGET TEAMS DISTRIBUTE private(idn,id,ind,iq,udu,edu,feq)
    do idn=0,size_fluid-1
       id=fluid_id(idn)
       udu=0.d0
       do ind=0,dim-1
          udu=udu+u(id*dim+ind)**2
       enddo

       do iq=0,nq-1
          edu=0.d0
          do ind=0,dim-1
             edu=edu+e(iq*dim+ind)*u(id*dim+ind)
          enddo

          feq=t(iq)*(p(id)*3.d0+rho0*(3*edu+4.5*edu**2-1.5*udu))
          fb(id*nq+iq)=(1-ome)*f(id*nq+iq)+ome*feq
       enddo
    enddo
    !$OMP END TARGET TEAMS DISTRIBUTE

endsubroutine Collision
\end{lstlisting}

In the \texttt{BoundaryCondition} subroutine, the communications among different nodes are required. In this case the host data need to be updated.
\begin{lstlisting}[style=fort]
subroutine BoundaryCondition
    integer idn,id,iq

    ! OpenMP and MPI synchronization before propagation:
    !$OMP TARGET UPDATE from(fb)
    call PassF(fb)
    !$OMP TARGET UPDATE to(fb)

    ...
endsubroutine BoundaryCondition
\end{lstlisting}

In the \texttt{Monitor} subroutine, user defined global variables are monitored. The calculation of global values usually requires both MPI reduction and OpenMP reduction.
\begin{lstlisting}[style=fort]
subroutine Monitor
    integer idn,id
    double precision um_global

    ! Find the MPI local maximum magnitude of velocity
    um=0
    ! OpenMP reduction
    !$OMP TARGET TEAMS DISTRIBUTE map(tofrom:um) private(idn,id) reduction(max:um)
    do idn=1,size_fluid-1
       id=fluid_id(idn)
       um=max(um,sqrt(u(id*dim)**2+u(id*dim+1)**2))
    enddo
    !$OMP END TARGET TEAMS DISTRIBUTE
    call MPI_BARRIER(CART_COMM,ierr)

    ! MPI reduction
    call MPI_REDUCE(um,um_global,1,MPI_DOUBLE_PRECISION,MPI_MAX,0,CART_COMM,ierr)
    ! Print
    if(rank.eq.0)then
       write(*,*)"T=",iter/t_intv
       write(*,*)"Umax=",um_global/(charlength*t_intv)
       write(*,*)"-------------------"
    endif
endsubroutine Monitor
\end{lstlisting}

\section{Example} \label{example}
\begin{figure}
\centering
\includegraphics[width=\textwidth]{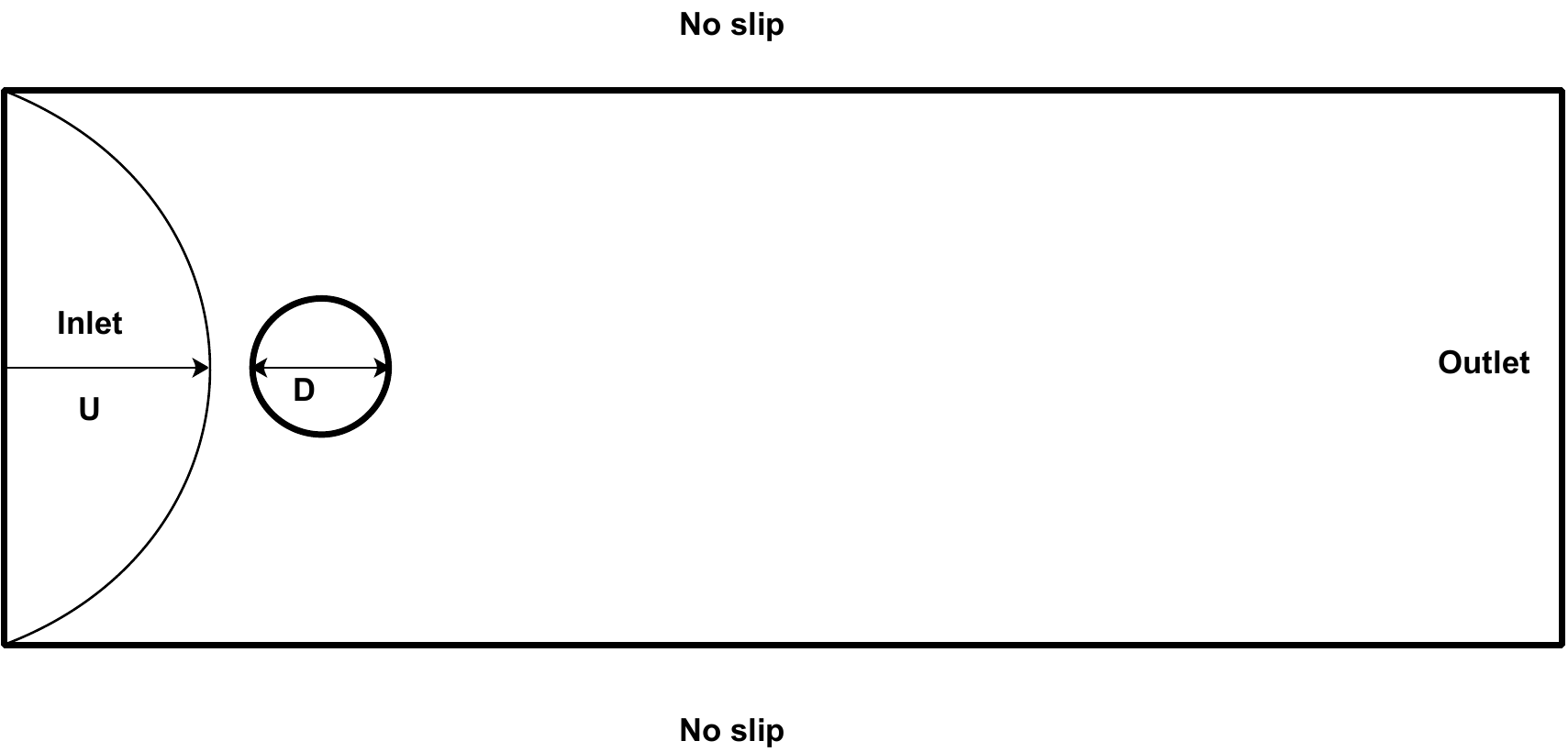}
\caption{Geometry setup of channel cylinder flow.}\label{geo}
\end{figure}

 To demonstrate that our code delivers qualitatively reasonable results, we perform a  $2$D simulation of flow past a cylinder flow. This test case has been a benchmark for numerous simulation codes and will allow us to qualitatively assess the validity of the code for CFD simulations. The flow configuration is a time-dependent flow through a channel with a circular obstacle.   The geometry is depicted by Fig. \ref{geo}. The domain is rectangular,  with top and bottom no-slip boundaries. The left side is an inlet with a parabolic velocity profile, and the right side is convective outflow. The cylinder center is on the centerline of the channel and positioned close to the inlet boundary so it begins to shed vortices soon after the start of the simulation. The geometry is non-dimensional with the domain normalized by the channel length.
 We consider two cases with this setup. The first one is a laminar flow. The channel width is 1, the distance from cylinder center to inlet is 0.5 and the characteristic length (cylinder diameter $D$) is 0.03125. The relaxation parameter ($1/(\tau/\delta_t+0.5)$) is chosen as 1.8, where the relaxation time $\tau$ is dependent to the kinematic viscosity $\nu$:
 \begin{eqnarray}
 \tau=\frac{\nu}{c_s^2}.
 \end{eqnarray}
The non-dimensional time is normalized by $D^2/\nu$. The Reynolds number ($UD/\nu$, where $U$ is the characteristic velocity in Fig. \ref{geo}) is 20.
 The length and width of the channel in lattice units are 1024 and 1024 respectively.

 The drag coefficient is calculated by \begin{eqnarray}
\frac{\vF}{\rho U^2 r},
\end{eqnarray}
where $\vF$ is the drag force, and $r$ is the cylinder radius.  The boundary forces are evaluated by momentum exchange method \cite{Ladd1994a,Ladd1994b,MomentumExchange}:
\begin{eqnarray}
\vF=\sum_{\substack{\vx\in\\\text{wall}\\\text{neighbor}}}\delta_t\sum_{\substack{\alpha\in\\\text{outgoing}\\\text{direction}}}\left[f^{\ast}_{\alpha}(\vx,t)+f_{\oalpha}(\vx-\ve_{\oalpha}\delta_t,t-\delta_t)\right]\ve_{\alpha}.
\end{eqnarray}
As the flow gets to steady state, the drag coefficient converges to 2.06, which is comparable to the reported data in \cite{liu2021}.

 In the second case, the channel width is 0.25, the distance from cylinder center to inlet is 0.125 and the characteristic length is 0.03125. The relaxation parameter is chosen as 1.953. The Reynolds number is 100.
 The length and width of the channel in lattice units are 512 and 128 respectively.

\begin{figure}
\centering
\includegraphics[width=\textwidth]{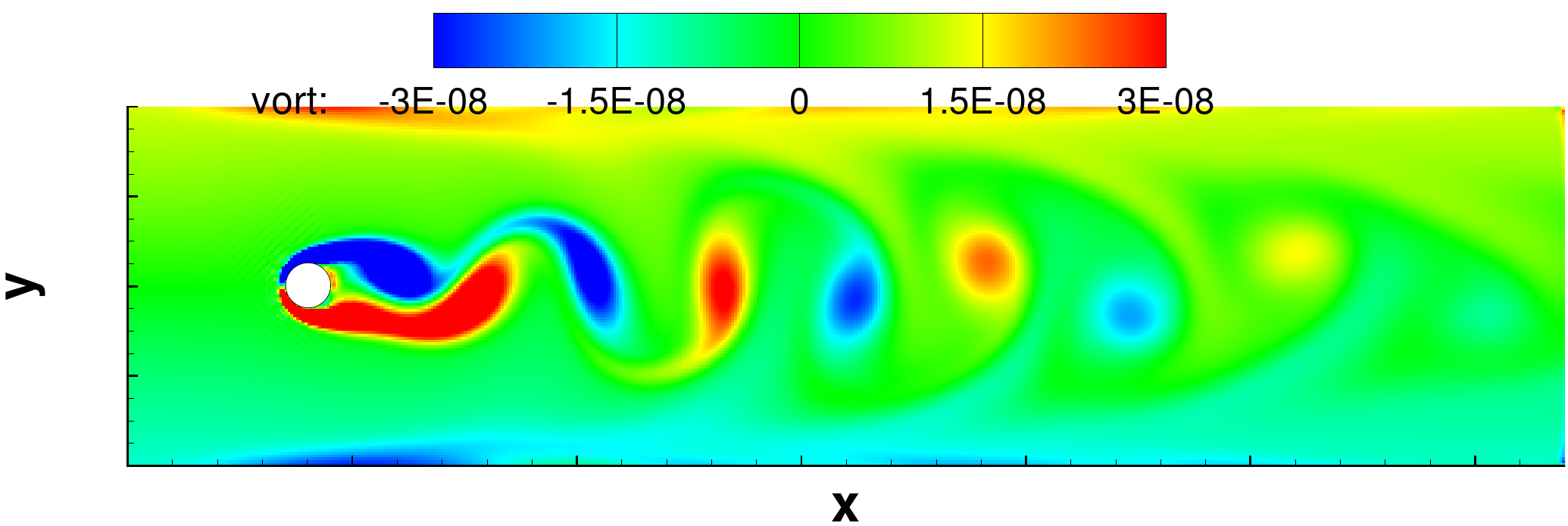}
\caption{Vorticity field of channel cylinder flow.}\label{vort}
\end{figure}

\begin{figure}
\centering
  \begin{subfigure}[b]{\textwidth}
    \centering
    \includegraphics[width=0.8\textwidth]{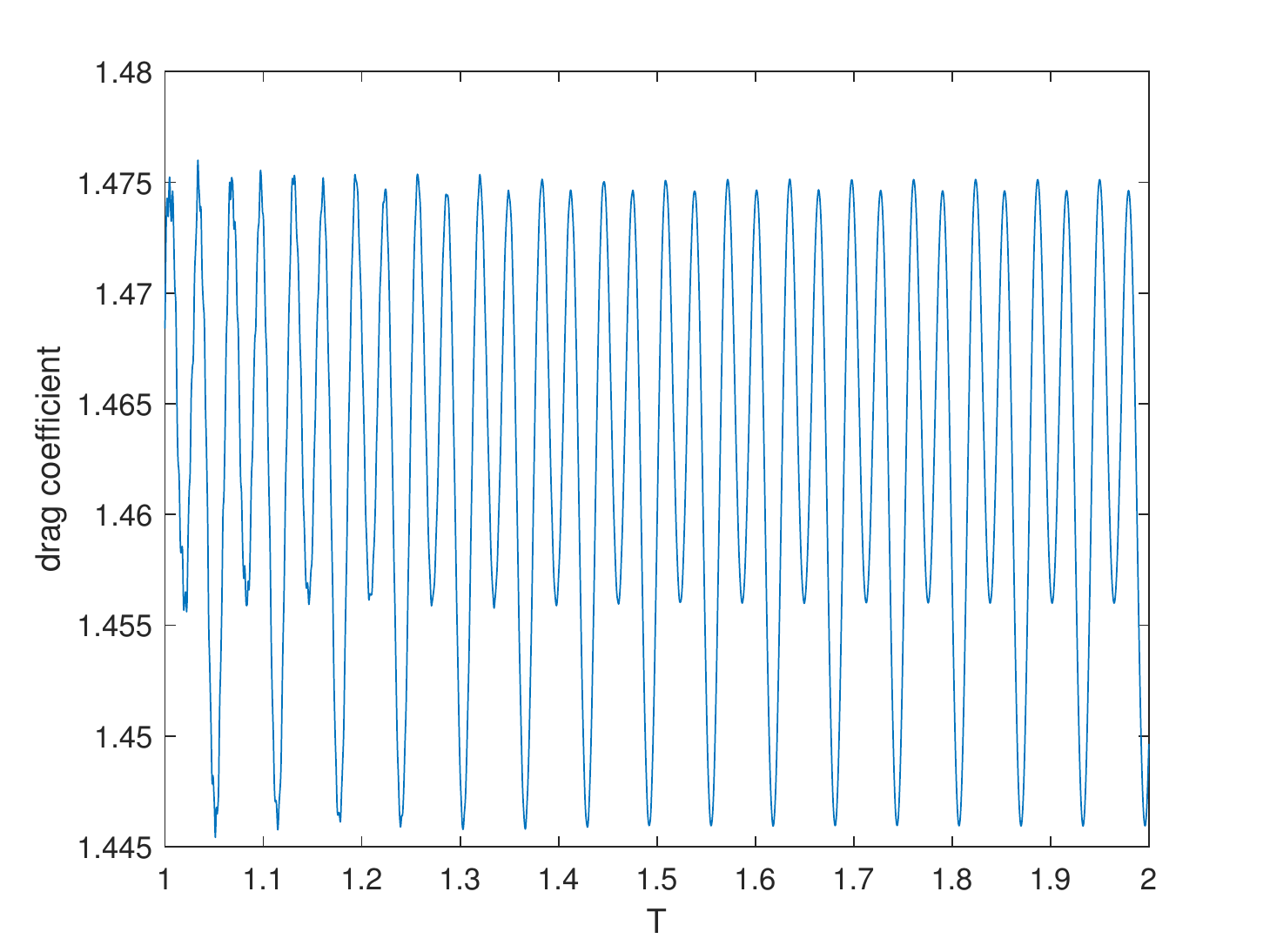}
    \caption{History of drag coefficient.}
  \end{subfigure}
  \\
  \begin{subfigure}[b]{\textwidth}
    \centering
    \includegraphics[width=0.8\textwidth]{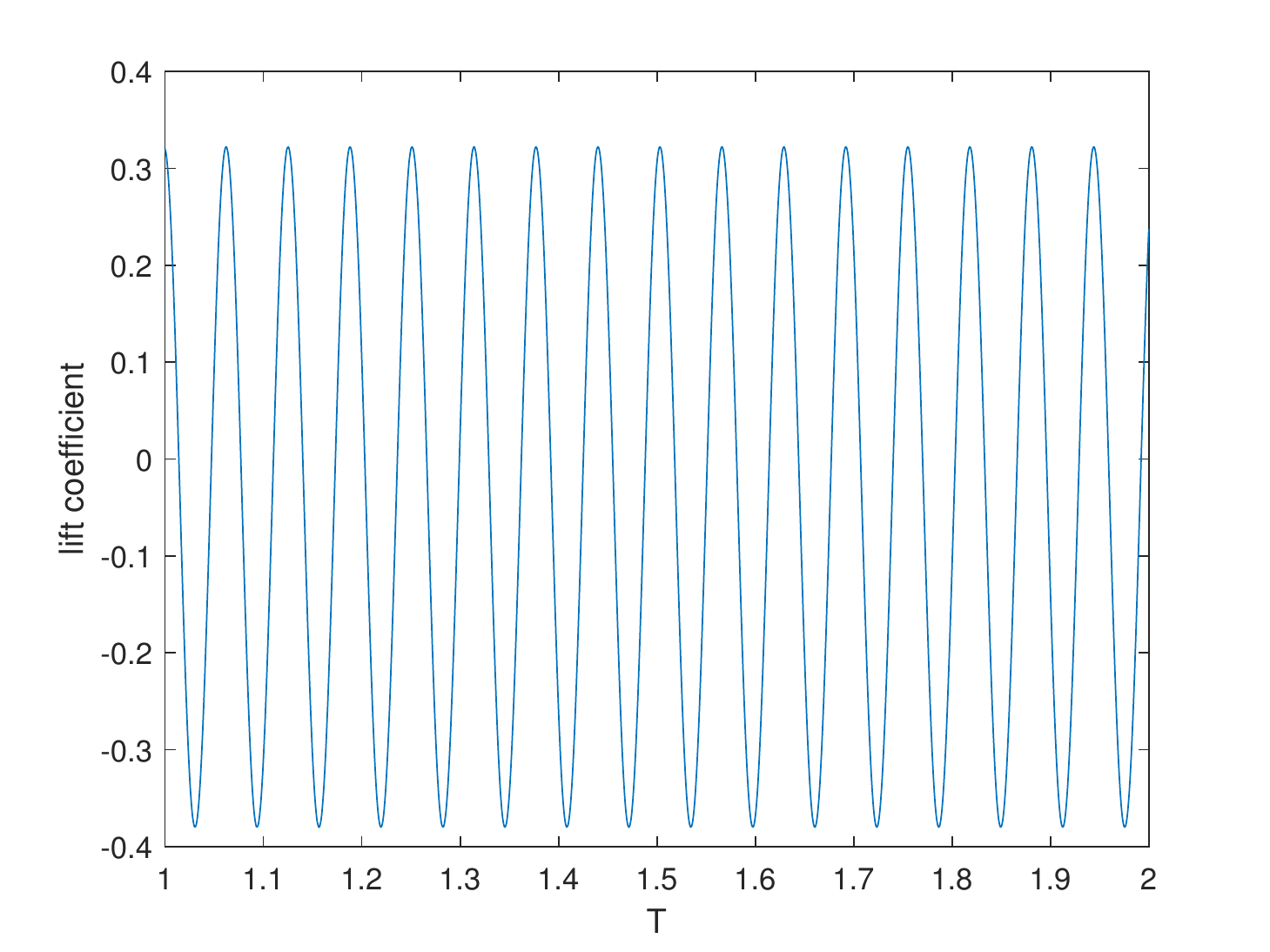}
    \caption{History of lift coefficient.}
  \end{subfigure}
\caption{Histories of drag and lift coefficients for 2D cylinder flow at Reynolds number 100.}\label{coef}
\end{figure}

The initial state of flow is a steady-state inviscid flow with randomly distributed perturbation. As the normalized time goes from 0 to 0.5, the periodic shedding of vortices appear as shown in Fig. \ref{vort}. The vortex street is named after the famous scientist Theodore von Kármán, and is expected when Reynolds number is larger than 47. The histories of drag and lift coefficients are shown in Fig. \ref{coef}.
 The drag coefficient oscillates around 1.46.

The simulations are implemented on ThetaGPU (1 node, 8 ranks) and accelerated by 8 GPUs.

\section{Validation of GPU Offloading}\label{validation}

\begin{figure}
\centering
\includegraphics[width=0.5\textwidth]{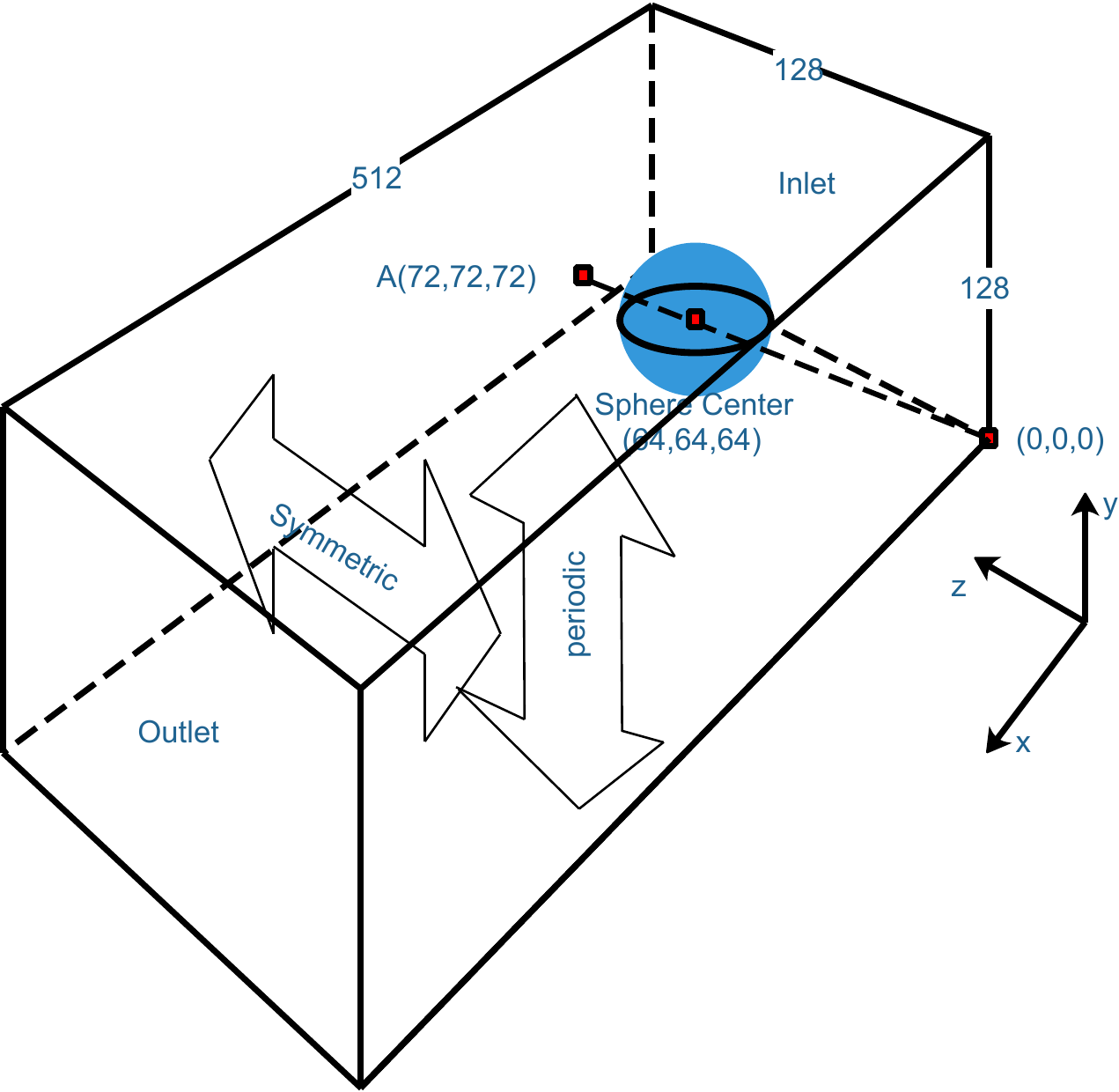}
\caption{Geometry setup of sphere flow.}\label{pointAsketch}
\end{figure}

The GPU acceleration is validated by a flow past a 3D sphere problem. The geometry of the problem is shown in Fig. \ref{pointAsketch}. The cuboid is 512 by 128 by 128 in lattice units. The back surface is inlet and the front surface is outlet, the left and right boundaries are symmetric, and the top and bottom boundaries are periodic. The radius of the sphere is 8, and the sphere center is at (64,64,64). The DVM in the code is D3Q27. This application is also used to test the performance of the code in the next section.

\begin{figure}
\centering
  \begin{subfigure}[b]{0.45\textwidth}
    \centering
    \includegraphics[width=\textwidth]{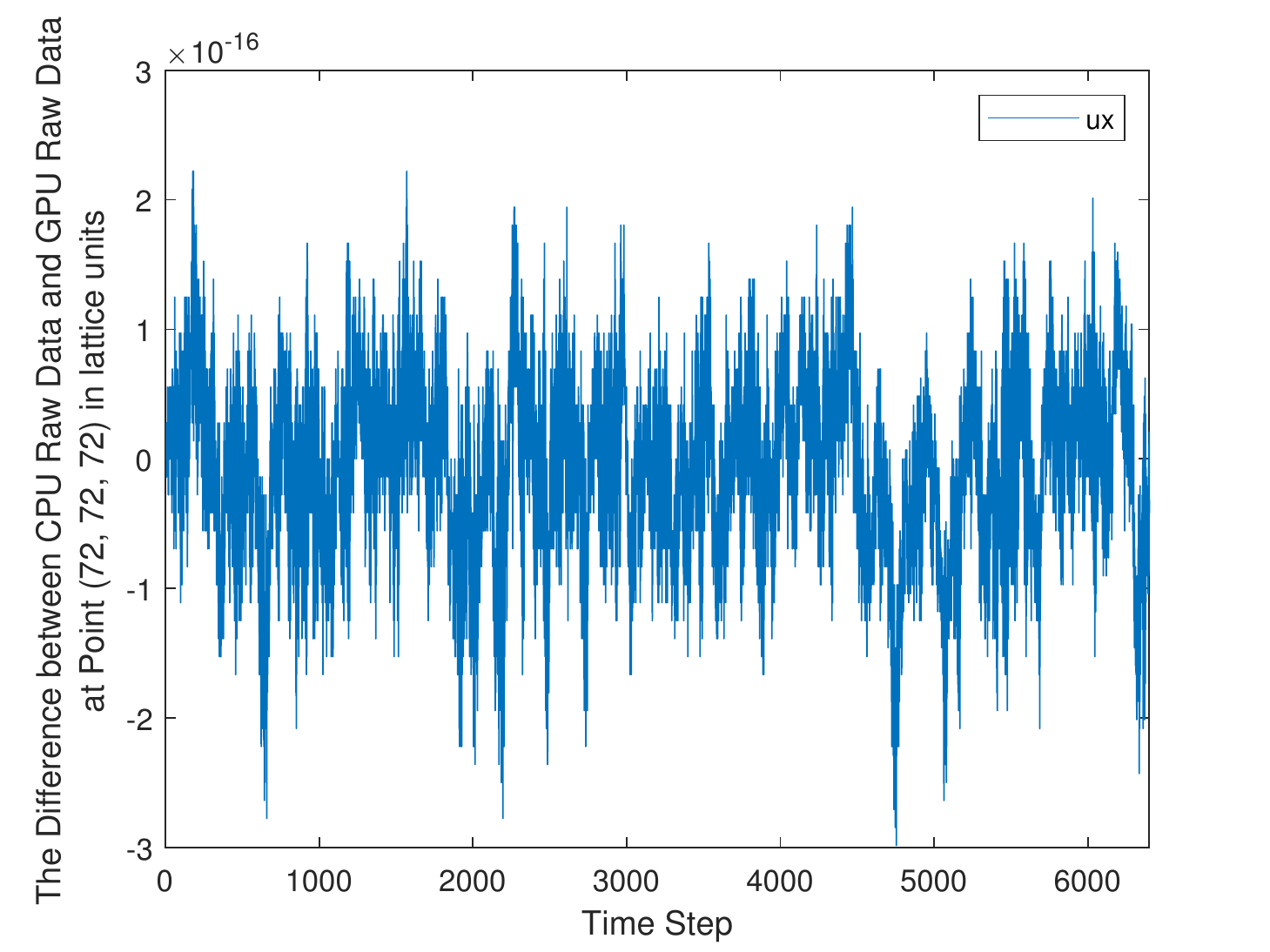}
    \caption{Difference of $u_x$ from CPU and GPU results.}
  \end{subfigure}
  \hfil
  \begin{subfigure}[b]{0.45\textwidth}
    \centering
    \includegraphics[width=\textwidth]{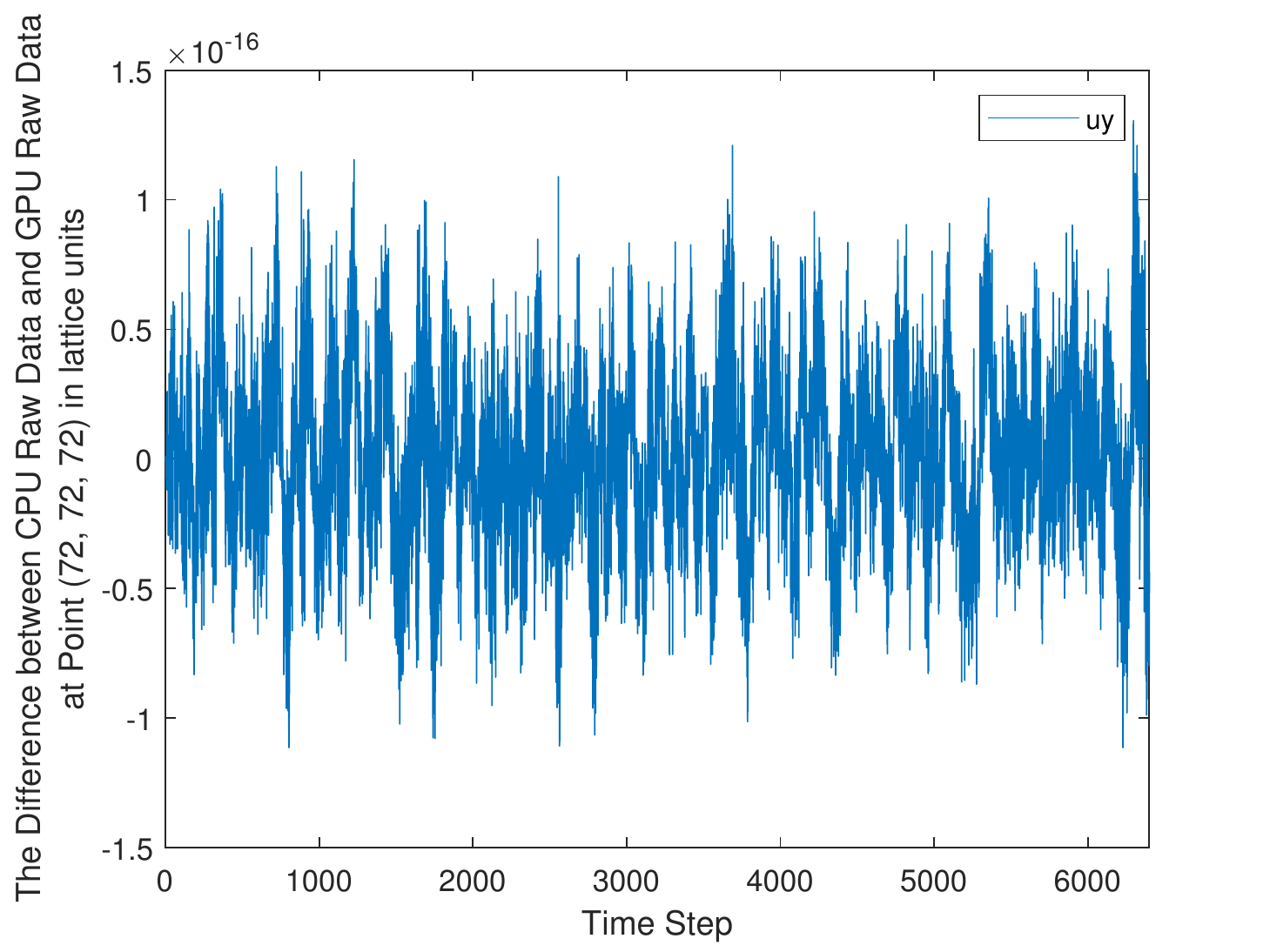}
    \caption{Difference of $u_y$ from CPU and GPU results.}
  \end{subfigure}\\
  \begin{subfigure}[b]{0.45\textwidth}
    \centering
    \includegraphics[width=\textwidth]{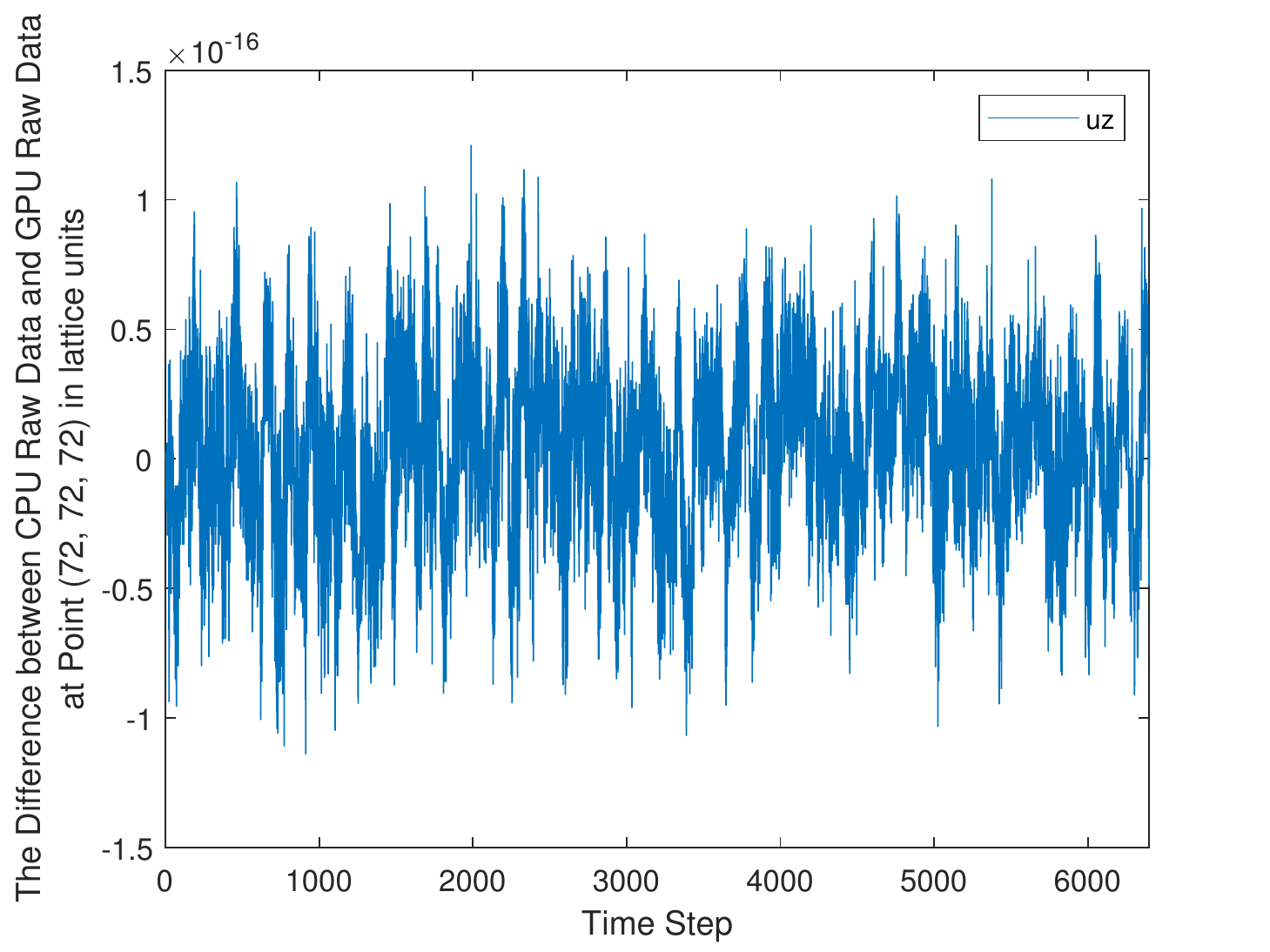}
    \caption{Difference of $u_z$ from CPU and GPU results.}
  \end{subfigure}
  \hfil
  \begin{subfigure}[b]{0.45\textwidth}
    \centering
    \includegraphics[width=\textwidth]{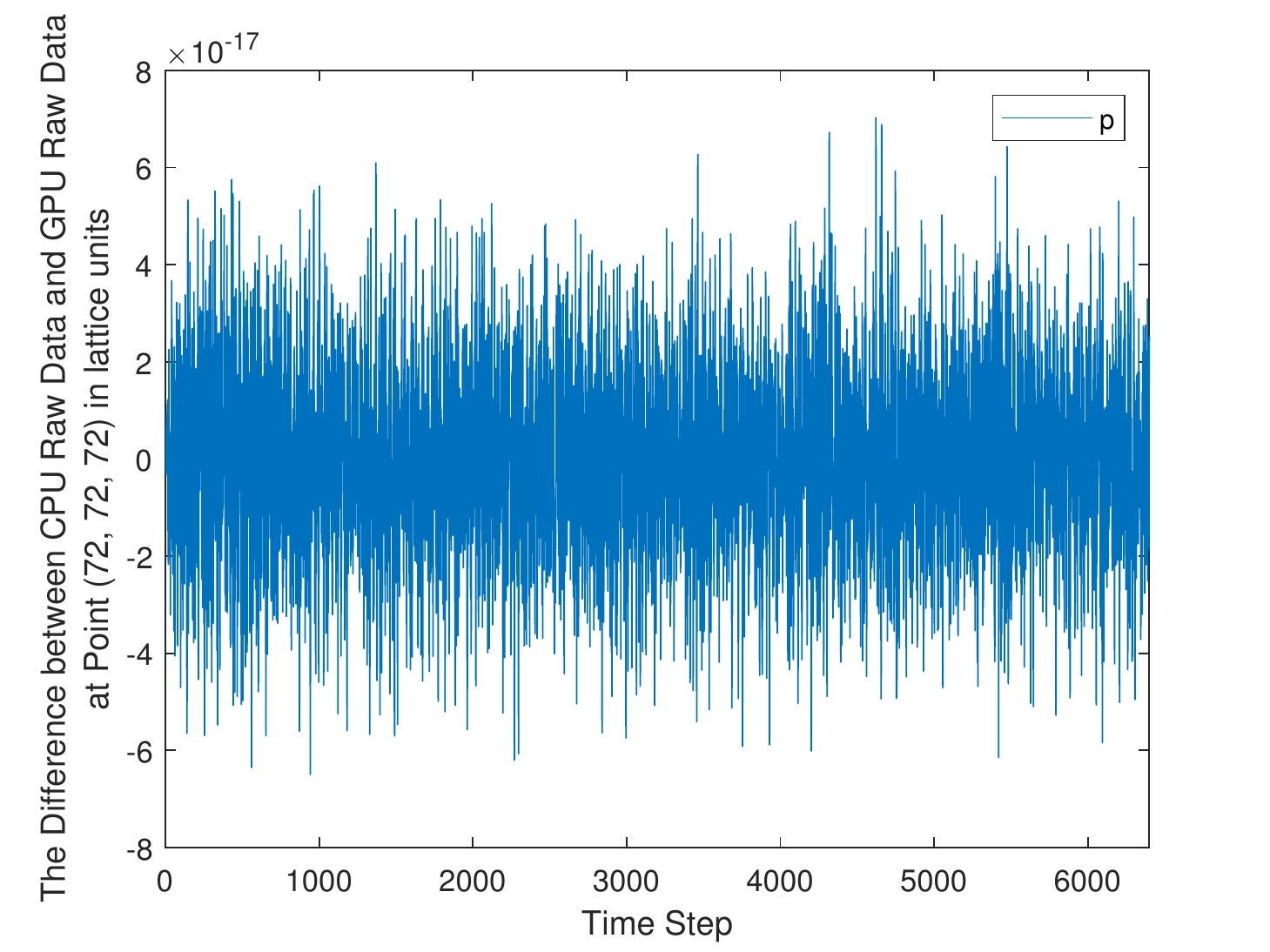}
    \caption{Difference of pressure from CPU and GPU results.}
  \end{subfigure}
  \caption{Raw data comparison between CPU and GPU cases. \label{Acomp}}
\end{figure}

Here we inspect the data at point A (72,72,72) in Fig. \ref{pointAsketch}. The simulation is run on ThetaGPU with CPU-only (8 MPI ranks) and with CPU + GPU offloading (8 A100 GPUs) using MPI and OpenMP. Additional details of a ThetaGPU node can be found in Table \ref{table: thetagpu}. The geometry of the problem is shown in Fig. \ref{pointAsketch}. The velocity and pressure at point A from the two cases are compared in Fig. \ref{Acomp}. As shown in the comparison, the differences between the two data are at the level of $10^{-16}$ or $10^{-17}$.

\begin{xltabular}{\linewidth}{ l | c }
  \caption{Specifications of ThetaGPU node}
 \label{table: thetagpu}\\
 \hline \hline

\textbf{\normalsize Component} & \textbf{\normalsize Description (Per Node)}  \\
 \hline
\endfirsthead
 \hline \hline

\textbf{\normalsize Component} & \textbf{\normalsize Description}  \\
 \hline
\endhead

\textbf{\texttt{cpu}} & AMD Rome 64-Cores ($\times$ 2) \\ \hline

\textbf{\texttt{memory}} & DDR4, 1 TB \\ \hline

\textbf{\texttt{gpu}} & NVIDIA A100 ($\times$ 8) \\ \hline

\textbf{\texttt{gpu-memory}} & 320 GB \\ \hline

\end{xltabular}

\section{Performance}\label{performance}
\begin{figure}
\centering
  \begin{subfigure}[b]{\textwidth}
    \centering
    \includegraphics[width=\textwidth]{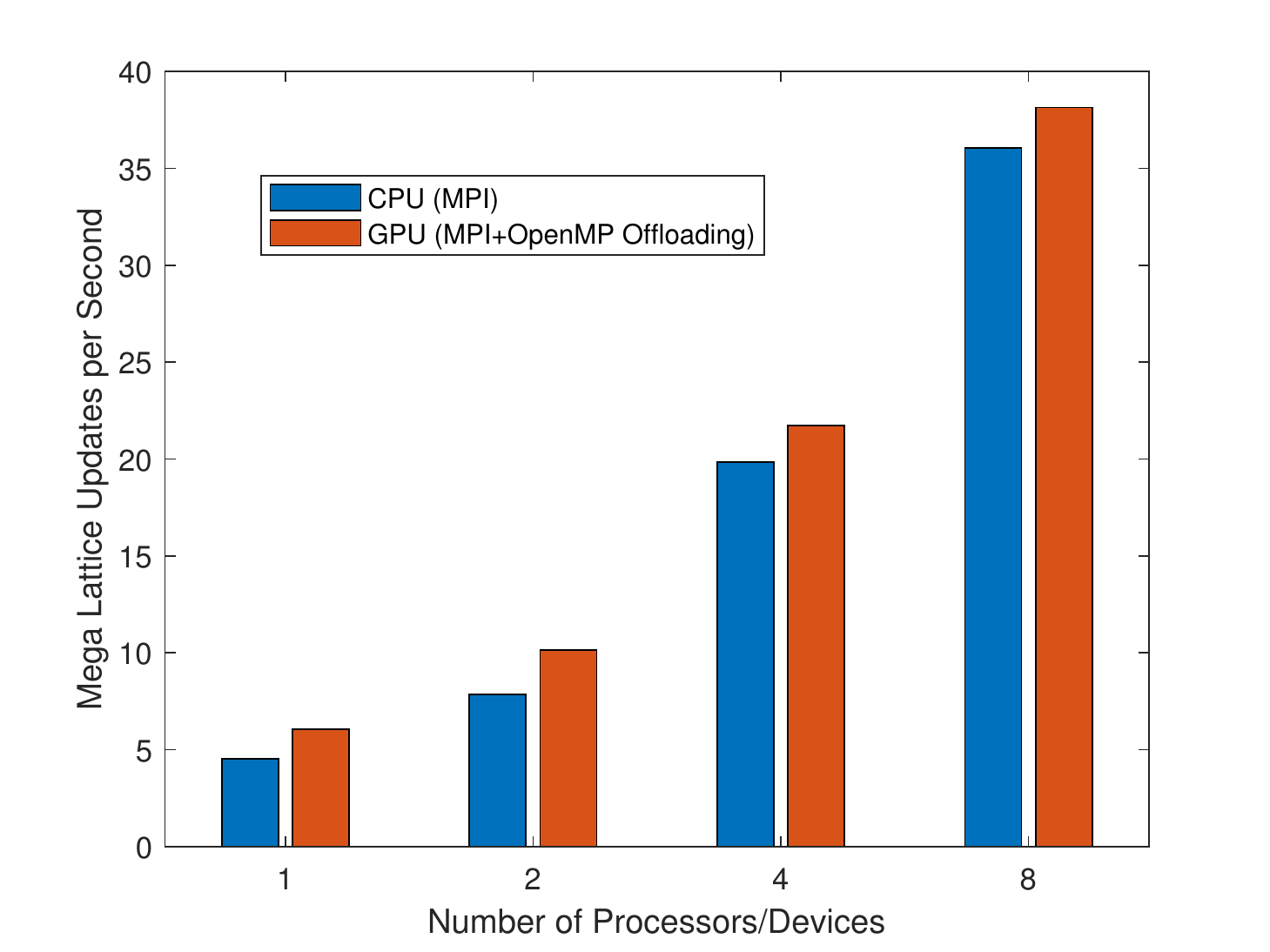}
    \caption{Mega Lattice Updates per Second.}\label{persecond}
  \end{subfigure}
  \\
  \begin{subfigure}[b]{\textwidth}
    \centering
    \includegraphics[width=\textwidth]{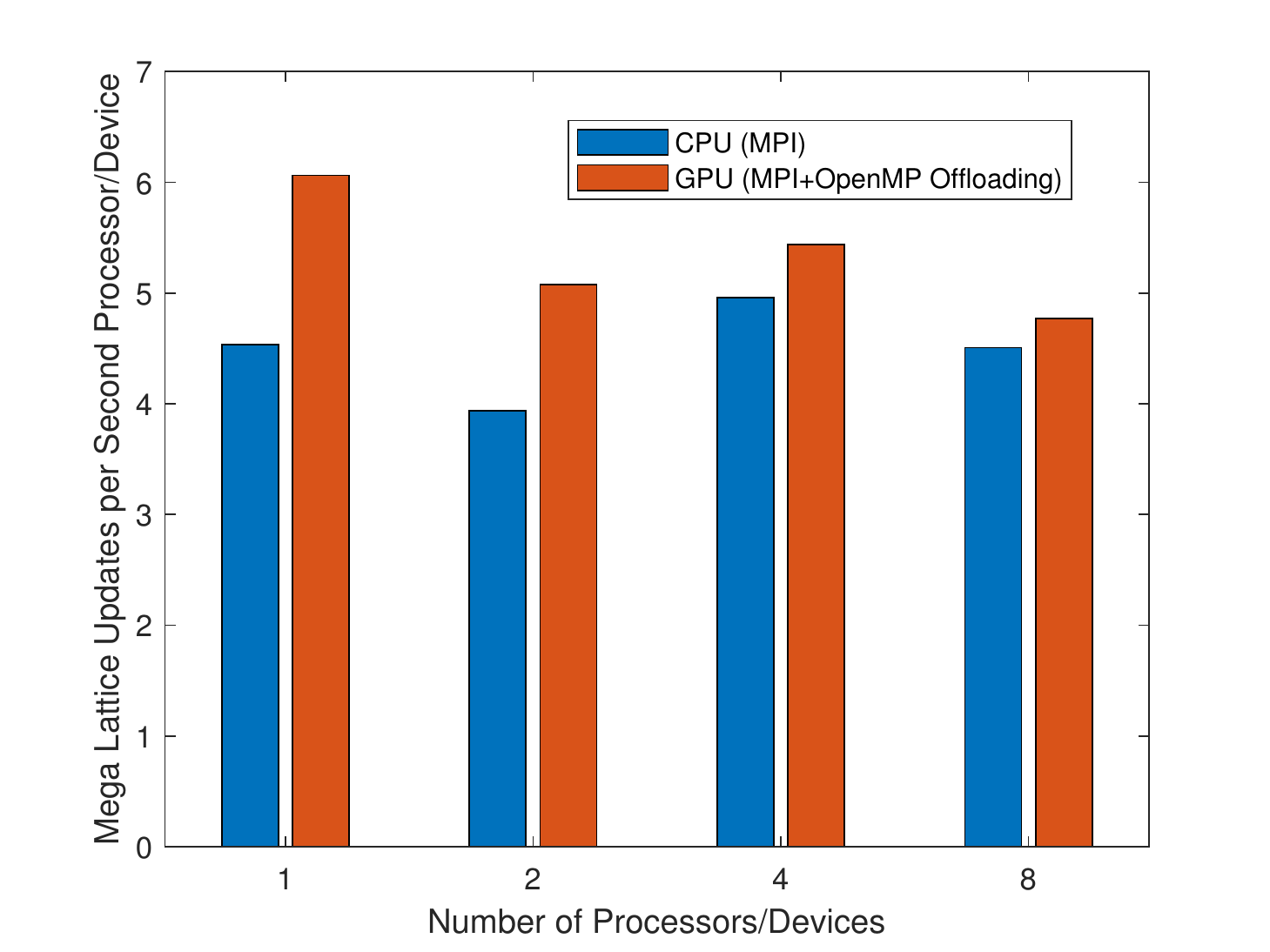}
    \caption{Mega Lattice Updates per Second per Processor/Device.}\label{persecondproc}
  \end{subfigure}
  \caption{Performance comparison of CPU and GPU cases. \label{performance}}
\end{figure}

The problem in the above section is employed to test performance of the code. The code is run for 100 time steps. The simulation is implemented with 1, 2, 4 and 8 CPUs and 1, 2, 4 and 8 GPUs respectively. The performances are evaluated by Mega Lattice Updates per Second (MLUPs). This value $m$ can be expressed by:
\begin{eqnarray}
m=\frac{\text{lattice points in the whole domain} \times \text{ iterations}}{\text{computing time} \times 10^6}.
\end{eqnarray}
The results are shown in Fig. \ref{performance}. The geometry of the problem is shown in Fig. \ref{pointAsketch} Fig. \ref{persecond} shows the code scales properly in both CPU and GPU cases. The GPU cases have slightly better performances than the CPU cases. Fig. \ref{persecondproc} shows the processor efficiencies. In the CPU case, 1 processor has the highest efficiency. In the GPU case, 4-GPU acceleration is more efficient.

\section{Conclusion}\label{conclusion}
In this work, we outline and give details for an ECP proxy app called \texttt{IMEXLBM}, an open-source, self-contained code unit, with minimal dependencies, that is capable of running on heterogeneous platforms. We have demonstrated functionality of the code by reviewing results from a 2D simulation of flow past a circle. Results from this simulation were presented in the form of a visualization which show the expected physics. Further work is required to show quantitative accuracy by investigating convergence. A thorough description of the Fortran 90 codes is also provided where we show critical subroutines and OpenMP pragmas are inserted to enable offloading to the GPU. Results generated by the GPU are validated against CPU-only calculations with differences that have magnitudes as small as $10^{-17}$.  Finally performance of the code was investigated on a single node of the ThetaGPU cluster. For a fixed problem we see that the code does strong scale in both scenarios: CPU-only as well as GPU-enabled calculations. Further investigation is required to understand performance. In particular, a roofline analysis will help to assess we can do any further analysis to improve performance.

\section*{Acknowledgement}
This research was supported by the Exascale Computing Project (ECP), Project Number: 17-SC-20-SC, a collaborative effort of two U.S. Department of Energy organizations—the Office of Science and the National Nuclear Security Administration—responsible for the planning and preparation of a capable exascale ecosystem—including software, applications, hardware, advanced system engineering, and early testbed platforms—to support the nation’s exascale computing imperative.

The research used resources of the Argonne Leadership Computing Facility, which is supported by the U.S. Department of Energy, Office of Science, under Contract DE-AC02-06CH11357.
\bibliographystyle{elsarticle-num}
\bibliography{imexlbm}

\end{document}